# MODELING SUBSTITUTION AND INDEL PROCESSES FOR AFLP MARKER EVOLUTION AND PHYLOGENETIC INFERENCE

By Ruiyan Luo[1] and Bret Larget[2]

*University of Wisconsin—Madison*

The amplified fragment length polymorphism (AFLP) method produces anonymous genetic markers from throughout a genome. We extend the nucleotide substitution model of AFLP evolution to additionally include insertion and deletion processes. The new Sub-ID model relaxes the common assumption that markers are independent and homologous. We build a Markov chain Monte Carlo methodology tailored for the Sub-ID model to implement a Bayesian approach to infer AFLP marker evolution. The method allows us to infer both the phylogenies and the subset of markers that are possibly homologous. In addition, we can infer the genome-wide relative rate of indels versus substitutions. In a case study with AFLP markers from sedges, a grass-like plant common in North America, we find that accounting for insertion and deletion makes a difference in phylogenetic inference. The inference of topologies is not sensitive to the prior settings and the Jukes–Cantor assumption for nucleotide substitution. The model for insertion and deletion we introduce has potential value in other phylogenetic applications.

**1. Introduction.** The amplified fragment-length polymorphism (AFLP) technique, first developed by Vos et al. (1995), is a powerful tool to produce DNA fingerprints of organismal genomes. The generation of AFLP markers begins by breaking whole genomic DNA into fragments, typically with two restriction enzymes. Double-stranded adaptors specific to each restriction enzyme attach to the end of each fragment, forming caps. A small fraction of the fragments, selected by a specific primer pair, are amplified using a polymerase chain reaction and separated by size using gel electrophoresis.

Received May 2007; revised September 2007.
[1]Supported by NIH Grants R01 GM068950-01 and R01 GM069801-01.
[2]Jointly appointed in the departments of Botany and of Statistics, University of Wisconsin–Madison. Supported by NSF Grant 0445453, NIH Grants R01 GM068950-01 and R01 GM069801-01.

Bands exhibiting variability among the separate individuals under study are the genetic markers. The resulting data are usually recorded as a 0/1 matrix—allele absent or allele present. For example, our case study of 14 sedges involves data at 126 AFLP markers [Supplementary Table 1 in Luo and Larget (2009)].

Because of their high replicability [Jones et al. (1997); Powell et al. (1996)], low cost, and ease of use, AFLP markers have emerged as an important genetic marker with broad applications. One increasingly common use of AFLP marker data is as a source of genetic information for phylogenetic inference, the estimation of evolutionary trees from genetic data. AFLP markers are less prone to homology problems than other anonymous DNA fragment methods such as randomly amplified polymorphic DNA fragments (RAPD) or inter-simple sequence repeat (ISSR) polymorphisms [Wolfe and Liston (1998)]. Moreover, as a multilocus method, AFLPs have the benefit of integrating phylogenetic signals from loci distributed throughout the genome, reducing the degree to which lineage sorting and reticulate evolution (hybridization) are expected to confound efforts to reconstruct phylogenies among rapidly radiating taxa [Albertson et al. (1996)]. Because of these qualities, AFLPs have come into increasingly frequent use in phylogenetic studies among closely-related species.

Binary genetic data, such as AFLP, have been analyzed by a simple two-state Markov model [Mau and Newton (1997)] as implemented in MrBayes [Huelsenbeck and Ronquist (2001)]. A more accurate approach models nucleotide substitutions within the AFLP marker itself [Luo, Hipp and Larget (2007)]. No procedures are yet available that accomodate insertion or deletion (indel) events, which are mutational processes that can affect AFLP markers. For example, indels can also result in sequence changes that affect AFLP markers. For example, indel processes could cause the loss of a marker by removing part of a restriction site or the neighboring amplification sites. Indel events in the interior region (Figure 1) have the potential to cause a single homologous locus to result in two or more markers of different lengths in different species. We call such a situation a locus-splitting. Furthermore, it is possible that markers with identical lengths could be produced by different loci. We call a marker that is produced by multiple loci a superposition. With the introduction of indel processes, the common assumptions in the substitution-only model of Luo, Hipp and Larget (2007) and the MrBayes method that each marker is associated with a single genetic locus and that the loci in different individuals corresponding to the same AFLP marker are homologous (derived from a single locus in a common ancestor) are invalidated. The model introduced here differs from several indel models previously described [Miklós, Lunter and Holmes (2004); Redelings and Suchard (2005); Thorne, Kishino and Felsenstein (1991, 1992)].

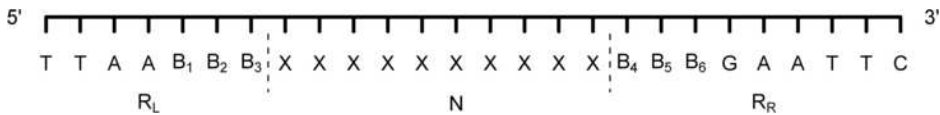

FIG. 1. *A typical fragment corresponding to an AFLP marker is partitioned into three parts. The first and third parts, referred to as the "left end region" and "right end region," respectively, must match specific sequences exactly to be restricted and amplified. The first four bases of the left end region are a restriction site and the bases $B_1B_2B_3$ correspond to three bases in one of the primers. The right end region has bases $B_4B_5B_6$ that correspond to three bases in the other primer followed by a six base restriction site. The sequence in the second part is called the "intermediate region." Throughout the paper, the number of bases in the left and right end regions are denoted as $R_L$, and $R_R$, respectively. $R = R_L + R_R$ is the total number of bases in the end regions. The number of bases in the intermediate region is denoted as $N$ or $N_{\text{node}}$ for a specific node.*

We describe a model that incorporates both substitution and indel processes, and we present a Bayesian approach to infer phylogenies from AFLP marker data. We call this model the Sub-ID model. We begin by briefly describing the substitution-only model in Luo, Hipp and Larget (2007), upon which we will model the indel processes (Section 2). Then we study how indel events affect AFLP markers by examining the six complete genomes (Section 3). We describe the Sub-ID model in Section 4 and illustrate the Bayesian structure for phylogenetic inference from AFLP data in Section 5. In Section 6 we implement a novel Markov chain Monte Carlo (MCMC) approach for phylogenetic inference. We study the sensitivity of the model to prior settings with a simulated data set in Section 7 and apply the methodology to analyze AFLPs from several taxa in *Carex* Section *Ovales*, a group of sedges common in North America, in Section 8.

**2. The substitution-only model.** This section summarizes the substitution-only model described in Luo, Hipp and Larget (2007) and introduces the notation that will be used in the Sub-ID model. We partition a particular fragment corresponding to an AFLP marker into three regions (Figure 1) according to the process of AFLP data production. In the original protocol [Vos et al. (1995)], (1) the two restriction enzymes used are *Eco*RI, which cleaves DNA whenever the sequence "GAATTC" appears in the $5'$ to $3'$ direction, and *Mse*I, which cleaves DNA at a four-base recognition sequence "TTAA"; (2) only the *Eco*RI adaptors are fluorescently labeled, so fragments flanked by two *Mse*I sites are invisible and do not form markers; (3) the primer pair, one for each adaptor, matches a portion of the corresponding adaptor and restriction site plus three additional bases. By design, only fragments whose sequence includes the specific extra three bases in each end region corresponding to the primers for each restriction site are amplified. In practice, a researcher can use multiple primer pairs in order to find additional AFLP markers. A set of AFLP markers found with a single primer pair is called a *plate*. Our model is based on this protocol, but it can be easily modified for other choices of restriction enzymes or if primers of different lengths are used.

The first and third parts of the partition in Figure 1 include bases necessary for each end of the fragment to be cut (restriction sites) plus three additional bases necessary for amplification (amplification sites). We call these two parts the *left end region* and *right end region* in our new model, and denote the number of bases in them as $R_L$ and $R_R$, respectively. Let $R$ be the total length of the end regions. Obviously, $R = R_L + R_R$. The second part is an *intermediate region*. If we denote the number of bases in this region as $N$, then the corresponding measured marker length is $N + 39$, where 39 counts the number of bases in each primer $(19 + 19)$ and an extra adenine (A) base appended to the $3'$ end of *Taq* DNA polymerase.

The substitution-only model of AFLP evolution rests on the following assumptions: (1) each AFLP marker is associated with a single genetic locus in each individual; (2) the loci in different individuals corresponding to the same AFLP marker are homologous; (3) loci associated with visible markers are mutually independent; (4) bands are appropriately scored as present or absent; (5) each locus is represented by a band that is flanked either by an *Mse*I and an *Eco*RI site (with prior probability 32/33) or by two *Eco*RI sites (with prior probability 1/33); (6) a band is present for an *Mse*I/*Eco*RI (or *Eco*RI/*Eco*RI) fragment if there are zero mismatches among the 16 (or 18) necessary bases and no restriction sites between the restriction sites corresponding to the fragment ends; and (7) all sites evolve independently with the same rate according to a Jukes–Cantor model [Jukes and Cantor (1969)]. The model assumes only nucleotide substitution as a mutational process. Then marker loss is due either to mutation in the end regions or by gain of a restriction site in the intermediate region. In particular, nucleotide substitution at the end regions causes either loss of one of the restriction sites at one end of the fragment merging it with a neighboring fragment or causes a change in the amplification sites causing the fragment not to be amplified. A nucleotide substitution in the intermediate region usually has no effect, but can create a new restriction site resulting in the marker fragment being broken into two smaller fragments. At time $t$, let $M(t)$ be the number of mismatches among the $R$ bases in the end regions, and let $Z(t)$ be the presence/absence of cutters (or restriction sites) among the $N$ middle bases. Then $M(t)$ itself is a continuous-time Markov process on the state space $0, 1, 2, \ldots, R$, and $Z(t)$ can be approximated by a two-state continuous-time Markov chain, with $Z(t) = 1$ indicating the presence of cutters in the intermediate region, and $Z(t) = 0$ indicating the absence. We ignore the specific infrequent cases such as substitution that could change an AFLP band length to some longer length if a pre-existing flanking restriction site of the same type existed, or substitution in the intermediate region that introduces a new restriction site and amplification site and changes an AFLP band length to some shorter length.

For the process $M(t)$, if there are $R = r$ bases, the probability of changing from $i$ mismatches to $j$ mismatches in time $t$ is the sum of a product of two binomial probabilities, summing over the number of matches that become mismatches [as in Felsenstein (1992)]:

$$P_{ij}^{(r)}(t) = \sum_{k=\max(0,i-j)}^{\min(i,r-j)} \left( \binom{r-i}{j-i+k} p^{j-i+k}(1-p)^{r-j-k} \right) \times \left( \binom{i}{k} \left(\frac{p}{3}\right)^k \left(1-\frac{p}{3}\right)^{i-k} \right), \tag{1}$$

where $p = \frac{3}{4}(1 - e^{-4/3ut})$, and $u$ is the rate of substitutions per unit time per site. Typically, we measure $t$ in units of the expected number of substitutions per site and $u = 1$. The stationary probability of $i$ mismatches among $r$ independent sites is $\pi_i^{(r)} = \binom{r}{i} \frac{3^i}{4^r}$. For the process $Z(t)$, the stationary probability of no cutters in a fragment with $N = n$ bases in the intermediate region is

$$\pi_0^{(Z)} = P(Z(0) = 0) \approx (1 - \tfrac{1}{4^4})^{n-4+1}(1 - \tfrac{1}{4^6})^{n-6+1}. \tag{2}$$

The infinitesimal rate of moving from zero to at least one cutter is

$$q_{01}^{(Z)} \approx \frac{4(n-4+1)u}{4^4 - 1} + \frac{6(n-6+1)u}{4^6 - 1}. \tag{3}$$

Equations (2) and (3) are sufficient to determine the approximate probability transition matrix for $\{Z(t)\}$:

$$P^{(Z)}(t) = \begin{pmatrix} \pi_0^{(Z)} + (1 - \pi_0^{(Z)})\eta(t) & (1 - \pi_0^{(Z)})(1 - \eta(t)) \\ \pi_0^{(Z)}(1 - \eta(t)) & 1 - \pi_0^{(Z)}(1 - \eta(t)) \end{pmatrix}, \tag{4}$$

where $\eta(t) = \exp(-q_{01}^{(Z)} ut/(1 - \pi_0^{(Z)}))$.

**3. Indel processes and AFLP marker data.** We first investigate the prevalence of superpositions in real data by examining several sequenced genomes *in silico*. The AFLP data to be analyzed in this paper are from sedges, a grass-like plant, with genome size around 200 Mb. Among the fully sequenced and completely assembled genomes of similar size, we obtained six genomes with sizes between 90 Mb and 500 Mb from the NCBI genomes database [NCBI (2007)] and from the FlyBase database [FlyBase (2007)]. *Arabidopsis thaliana* (land plant), *Caenorhabditis elegans* (roundworm), *Drosophila melanogaster* (fruit fly), *D. pseudoobscura* (fruit fly), *Oryza sativa* (rice) and *Populus trichocarpa* (black cottonwood) have sizes about 116 Mb, 97 Mb, 180 Mb, 125 Mb, 389 Mb and 485 Mb, respectively. Using the same enzymes and primers pairs from Section 8, we found the fragments that could be amplified and obtained their lengths. Examining these fragment lengths, we determined the number of superpositions and observable AFLP markers with distinct lengths. We list the results for four species in Table 1. On average, 7.48%, 5.56%, 3.05% and 25.06% of markers are superpositions, respectively. For the two species of *Drosphilia*, we find that 1.96% and 8.73% of markers from *D. melanogaster* and *D. pseudoobscura* are superpositions, respectively. Under the Jukes–Cantor model, for a genome with size $G$, we can calculate the expected number of fragments amplified in the third step of AFLP production with length between 11 and 586 as

$$(5) \qquad G \times \tfrac{17}{4^6} \times \tfrac{33}{289} \times \tfrac{1}{4^6} \times 0.864.$$

Table 1
*The number of AFLP markers and superpositions for four genomes using nine pairs of primers. "# fragments" counts the number of fragments that could be amplified in the third step of AFLP production, and "# observed markers" counts the number of fragments with distinct lengths. When a superposition involves markers produced by more than two loci, the sum of # observed markers and # superpositions is less than # fragments. Values in parentheses give the genome sizes and the expected number of fragments calculated by (5)*

|                    | A. thaliana (116 Mb, 12) |    |   |    |    |   |    |    |    | C. elegans (97 Mb, 10) |   |   |   |   |   |    |   |   |
|--------------------|---|----|---|----|----|---|----|----|----|---|---|---|---|---|---|----|---|---|
| # fragments        | 9 | 18 | 5 | 14 | 18 | 8 | 25 | 12 | 16 | 9 | 9 | 1 | 3 | 6 | 8 | 11 | 5 | 9 |
| # observed markers | 8 | 15 | 5 | 14 | 16 | 8 | 14 | 12 | 14 | 9 | 8 | 1 | 3 | 6 | 8 | 8  | 5 | 9 |
| # superpositions   | 1 | 3  | 0 | 0  | 1  | 0 | 3  | 0  | 1  | 0 | 1 | 0 | 0 | 0 | 0 | 3  | 0 | 0 |
|                    | O. sativa (389 Mb, 39) |    |   |    |    |    |    |    |    | P. trichocarpa (485 M, 48) |    |    |    |    |    |    |    |    |
| # fragments        | 36 | 24 | 7 | 20 | 36 | 23 | 15 | 14 | 14 | 55 | 48 | 10 | 27 | 45 | 13 | 23 | 14 | 23 |
| # observed markers | 35 | 23 | 7 | 18 | 33 | 23 | 14 | 14 | 14 | 12 | 9  | 9  | 25 | 32 | 12 | 18 | 14 | 21 |
| # superpositions   | 1  | 1  | 0 | 2  | 3  | 0  | 0  | 0  | 0  | 10 | 6  | 1  | 2  | 7  | 1  | 3  | 0  | 2  |

The average fragment length is $4^6/17$, so the product of the first two factors is the expected number of fragments after digesting the whole genome. The third factor $33/289$ is the proportion of fragments ending with *Mse*I/*Eco*RI or *Eco*RI/*Eco*RI cutters. The fourth one $1/4^6$ is the proportion of fragments that match the additional bases in the primers and will be amplified. The probability that the fragment length is between 11 and 586 is 0.864 when approximating the length with a geometric distribution with mean $17/4096$. Here 11 and 586 are the lower and upper bound of the lengths of the intermediate part for visible markers. The expected values are comparable with the observations for the genomes considered (Table 1).

We also sought evidence for locus-splitting, which would tend to be more prevalent in closely related species and when the indel rate is high relative to the substitution rate. To date, only one pair (*D. melanogaster* and *D. pseudoobscura*) of closely related species with genome sizes comparable to those in sedges have been sequenced and fully assembled. Examining *D. melanogaster* and *D. pseudoobscura*, we do not find evidence of locus-splitting. Nor do we find such evidence among some fully sequenced yeast genomes (*Debaryomyces hansenii*, *Eremothecium gossypii* and *Saccharomyces cerevisiae*), but this is not surprising given their small size (about 20 Mb). The lack of evidence of locus-splitting among completely sequenced genomes can be attributed to limitations in the available data. Locus-splitting could be a component of AFLP marker evolution among sedges and other closely related species. We have obtained evidence of superpositions. Phylogenetic inference based on a substitution-only model ignoring insertion and deletions processes could be misleading since the assumption that each marker is associated with a single genetic locus and that the loci in different individuals corresponding to the same AFLP marker are homologous is invalidated. This motivates us to build a model incorporating both substitution and indel processes.

**4. Sub-ID model.** By incorporating indel processes, we relax the first two assumptions in Luo, Hipp and Largent (2007) but retain assumptions (3)–(7). Following the word usage in Thorne, Kishino and Felsenstein (1991), we refer to positions between sites as *links*. We assume that insertion and deletion happen at any links between bases equally likely with different rates and any number of bases could be inserted/deleted. We first describe the model in a single edge and then extend it to the whole tree. The substitution process in the Sub-ID model is the same as that in the substitution-only model introduced by Luo, Hipp and Largent (2007) and summarized in Section 2.

4.1. *Sub-ID model in a single lineage.* The model assumes a DNA sequence of infinite length, and indel events happen at any positions equally likely. At any link between two bases, insertion and deletion happen independently in accordance with Poisson processes with rates $\lambda$ and $\mu$, respectively, and the length of the inserted or deleted segment follows a geometric distribution on $1, 2, \ldots$, with mean $1/r$. Although the positions of indels are equally likely along the whole sequence, for the fragment corresponding to a particular marker, only the insertions within the fragment or deletions removing at least one base of the fragment can affect the presence/absence of the marker. Table 2 lists the events in our model. We call an indel event that destroys the end regions a *killing* event (including an insertion/deletion starting within a restriction site or a neighboring amplification site, and a deletion that removes one or more residues in the end regions). When a killing event occurs, we say that the fragment is killed. In this modeling, we neglect a few indel events that can affect AFLP marker data, as they are fairly improbable. For example, a very small proportion of possible indels within the end regions would leave the restriction site and amplification sequence intact. We ignore this possibility and the possibility that subsequent indels or substitutions could cause a killed restriction site to recover. Second, when an indel kills the restriction site itself, a new fragment would be formed by extending the fragment to the next restriction site. It is, however, very unlikely that the amplification site of the adjacent restriction site would match the primer, so we ignore this possibility. In addition, we ignore the possibility that indels within the intermediate region could add a new restriction site.

TABLE 2
*Description of the insertion-deletion events in the Sub-ID model*

| $w$ value | Insertion and deletion events modeled |
|---|---|
| $-1_R$ | Deletion starting within the end regions |
| $-1_P$ | Deletion starting before the fragment and removing $\geq 1$ base in the left end region |
| $-1_N$ | Deletion starting within the intermediate region and removing $\geq 1$ base in the right end region |
| $-1$ | Deletion starting within the intermediate region and not removing any bases in the right end region |
| $1_R$ | Insertion starting within the end regions |
| $1$ | Insertion starting within the intermediate region |

Now consider a marker with $N$ residues in the intermediate region, and $(R_L, R_R)$ bases at the left and right end regions, respectively. There are $R_L + R_R + N - 1$ possible starting positions for insertions, and $R_L + R_R + N$ possible starting positions in the fragment for deletions to affect the marker. In addition, long deletions before the fragment removing part of the left end region will destroy it and cause the marker to be absent. If we denote the link immediately before the fragment as position 0, the link that is $i$ bases after position 0 as position $i$, and the link that is $i$ bases before the fragment as position $-i$ ($i = 1, 2, \ldots$), then the deletions at position $-i$ with length greater than $i$ will remove part of the left end region of the marker and hence kill it. The rate of all such long deletions before the fragment that will kill the marker is the sum of rates over all negative positions[3]:

$$(6) \qquad \sum_{i=-1}^{-\infty} \sum_{j=-i+1}^{\infty} r(1-r)^{j-1} = \frac{1-r}{r}.$$

---

[3]The base counts between markers are finite, but sufficiently large such that approximating the sums in (6) as infinite makes no difference.

Under the assumption that insertion and deletion happen independently in accordance with Poisson rates $\lambda$ and $\mu$, respectively, the occurrence of an indel event that will affect the fragment is a Poisson process with rate

$$\eta = (R_L + R_R + N - 1)\lambda + (R_L + R_R + N + (1-r)/r)\mu, \tag{7}$$

where $(N+1) \times \lambda$ is the rate of insertions that start within the intermediate region and hence change the sequence size, $(R_L + R_R - 2) \times \lambda$ is the rate of insertions that kill the fragment by starting within the end regions, $(1-r)/r \times \mu$ is the rate of long deletions that start before the fragment and remove at least one base in the left end region, $N \times \mu$ is the rate of deletions which start within the intermediate region and either kill the fragment by removing one or more bases in the right end region or just change the sequence size, and $(R_L + R_R) \times \mu$ is the rate of deletions that start within the end regions and hence kill the fragment.

For an edge in a phylogenetic tree, let $h$ refer to the indel history along it. The indel history includes a sequence of events which are characterized by *time*, *type*, *position* and *length*. *Time* ($t$) describes when the event happens by regarding the time of the parent node as 0. *Type* ($w$, described in Table 2) indicates whether the event is an insertion or deletion, and whether it kills the fragment or not. *Position* ($s$) refers to the starting point of an event. *Length* ($l$) gives the number of bases inserted or deleted. The events are ordered by their occurrence times.

For an indel history with $k$ events, denote the characteristics of the $i$th event as time $t_i$, position $s_i$, type $w_i$ and length $l_i$. Let $N_i$ be the number of bases in the intermediate region after the $i$th event. Then $N_i = N_{i-1} + w_i \times l_i$ if $w_i = 1$ or $-1$. For convenience, if there are $k$ indel events during a period of time $T$, let $t_{k+1} = T$ and $N_{k+1} = N_k$. Let $N_0$ denote the length of the intermediate region before the first indel event. Let kill $= 1$ or 0 indicate the possible occurrence of a killing event.

Under these assumptions, the likelihood for an indel history $h = (t_i, w_i, s_i, l_i, i = 1, \ldots, k)$ is

$$p(h|N_0, R_L, R_R) = \exp\left(-\sum_{i=1}^{k+1} \eta_i \times (t_i - t_{i-1})\right) \\ \times \prod_{i=1}^{k}\{\lambda^{I(w_i>0)}\mu^{I(w_i<0)}r(1-r)^{(l_i-1)}\} \tag{8}$$

if there is no killing event, and

$$p(h|N_0, R_L, R_R) = \exp\left(-\sum_{i=1}^{k} \eta_i \times (t_i - t_{i-1})\right) \tag{9}$$

$$\times \prod_{i=1}^{k} \{\lambda^{I(w_i>0)} \mu^{I(w_i<0)} r(1-r)^{(l_i-1)}\}$$

if a killing event occurs, where $\eta_i = (R_L + R_R + N_{i-1} - 1)\lambda + (R_L + R_R + N_{i-1} + \frac{1-r}{r})\mu$. Equation (8) has one more term in the argument of the exponential function than does (9), which accounts for the likelihood that no indel events happen during the time period $(t_k, t_{k+1})$.

Consider a particular edge with length $T$. Let $h$ be the indel history along it, $(M_P, M_C)$ be the number of mismatches, and $(Z_P, Z_C)$ be the presence of cutters at the parent and child nodes, respectively. Let $N_P$ be the number of residues in the intermediate region of the fragment at the parent node. Given the lengths of the end regions $(R_L, R_R)$, the likelihood of indel and substitution histories $(h, M_C, Z_C)$ given the parent information $M_P, Z_P, N_P$ is

$$\begin{aligned}
(10) \quad & p(h, M_C, Z_C | T, M_P, Z_P, R_L, R_R, N_P) \\
& = p(h | N_P, R_L, R_R) \\
& \quad \times (I(\text{kill}=0) P^{(M)}_{M_P, M_C}(T|R) P^{(Z|h)}_{Z_P, Z_C}(T) + I(\text{kill}=1)),
\end{aligned}$$

where $P^{(M)}_{M_P, M_C}(T|R)$ is the transition probability for the number of mismatches changing from $M_P$ to $M_C$ during time $T$, and $P^{(Z|h)}_{Z_P, Z_C}(T)$ is the transition probability for the presence of cutters changing from $Z_P$ to $Z_C$ during time $T$.

4.2. *Modeling of AFLPs in a tree.* A rooted binary tree is used to describe the evolutionary history of AFLP markers. Our likelihood calculation requires a prior distribution for the state at the root. To account for the fact that loci producing markers are very atypical in the genome relative to random DNA segments that do not produce markers, we assume that each locus that produces one or more markers in the data would have produced an AFLP marker at some time in the lineage ancestral to the common ancestor of all taxa. We do not assume this time to be the same for each locus. So for each locus, we attach an edge from the root to an ancestor $\mathcal{A}$ with a locus-specific length and assume that at node $\mathcal{A}$ the process begins with zero mismatches and without cutters in the intermediate region. Then the number of edges is $\mathcal{E} = 2\mathcal{T} - 1$, and the number of nodes is $\mathcal{N} = 2\mathcal{T}$, where $\mathcal{T}$ is the number of taxa. We assume that the number of loci is random and that all loci share the same rooted tree topology. We also assume that fragments at different loci evolve independently.

Given an evolutionary history with $K$ loci, we specify a $K \times \mathcal{T} \times 2$ matrix $Y$ to describe the markers produced by each locus. Elements $y_{ki1}$ and $y_{ki2}$ ($k = 1, 2, \ldots, K$, $i = 1, 2, \ldots, \mathcal{T}$) denote the AFLP value and fragment length for the $i$th taxon at the $k$th locus, respectively. There are three possible values for $y_{ki1}$: 1, 0 and $-1$. Element $y_{ki1} = 1$ indicates that the $i$th taxon produces a marker with length $y_{ki2}$ at the $k$th locus. When $y_{ki1} = 0$, it indicates that the $i$th taxon retains the potential to produce a marker at the $k$th locus, but does not produce a marker due to either mismatches in the end regions or cutters in the intermediate region. And $y_{ki1} = -1$ represents that there is a killing event in the evolutionary history of taxon $i$ at locus $k$. If we let $x_{ij}$ denote the observed AFLP marker value for the $i$th taxon and the $j$th band, $x_{ij} = 1$ (or 0) indicates the presence (or absence) of the $j$th marker in the $i$th taxon, then there could be multiple three dimensional $K \times \mathcal{T} \times 2$ ($K$ could be different) matrices $Y$ that match the observed AFLP data $X$. For example, (11) and (12) are two assignments for the data set $X$ in Table 3 with 2 and 3 loci, respectively, where $Y_{k..}$ represents the assigned AFLP values and lengths for all taxa in the $k$th locus:

$$(11) \quad Y_{1..} = \begin{bmatrix} 1 & 50 \\ 1 & 51 \\ 1 & 50 \end{bmatrix}, \quad Y_{2..} = \begin{bmatrix} 1 & 51 \\ 1 & 52 \\ 0 & 51 \end{bmatrix}.$$

$$(12) \quad Y_{1..} = \begin{bmatrix} 1 & 50 \\ 1 & 51 \\ 1 & 50 \end{bmatrix}, \quad Y_{2..} = \begin{bmatrix} 1 & 51 \\ 1 & 51 \\ 0 & 51 \end{bmatrix}, \quad Y_{3..} = \begin{bmatrix} 1 & 51 \\ 1 & 52 \\ 0 & 51 \end{bmatrix}.$$

Table 3
*An example of observable AFLP data*

|         | 50 | 51 | 52 |
|---------|----|----|----|
| Taxon 1 | 1  | 1  | 0  |
| Taxon 2 | 0  | 1  | 1  |
| Taxon 3 | 1  | 0  | 0  |

Before calculating the likelihood of the indel history and substitution history for a tree, we mention a difference between the substitution-only model in Luo, Hipp and Larget (2007) and this Sub-ID model in terms of the distribution of number of mismatches and presence of cutters at the root of a tree. Luo, Hipp and Larget (2007) assume stationary distributions for these attributes at the tree root, but the Sub-ID model does not. The Sub-ID model, for every locus, assumes an ancestral node $\mathcal{A}$ for the root and this node $\mathcal{A}$ contains a fragment with the corresponding end regions for the locus, where there are no mismatches in the end regions and no cutters in the intermediate region. Hence, no killing event ever happens along the ancestral edges of this node. Thus, one more edge from the root, together with node $\mathcal{A}$, is attached in the tree under the Sub-ID model. The attached edge is assumed to have an exponential length a priori. Indel and substitution events happen along this edge, and hence determine the length of the fragment, the number of mismatches and the presence of cutters for the root, whose distributions are not stationary. We introduce this change in the model in part to account for the selection bias of considering only genomic loci that produce markers in some of the taxa of interest and also because modeling killing events does not leave a stationary distribution.

4.3. *Likelihood calculation.* Given the number of loci $K$, let $\mathbf{h_{id}} = (h_{\text{id}}^{(1)}, h_{\text{id}}^{(2)}, \ldots, h_{\text{id}}^{(K)})$ and $\mathbf{h_{sub}} = (h_{\text{sub}}^{(1)}, h_{\text{sub}}^{(2)}, \ldots, h_{\text{sub}}^{(K)})$ refer to the indel history and substitution history, respectively, over all loci. For any particular locus $k$, $h_{\text{id}}^{(k)} = (h_{id,e_1}^{(k)}, h_{id,e_2}^{(k)}, \ldots, h_{id,e_{\mathcal{E}}}^{(k)})$ denotes the indel history over all edges of the tree, and $h_{\text{sub}}^{(k)} = (M_0^{(k)}, M_1^{(k)}, \ldots, M_{\mathcal{N}}^{(k)}, Z_0^{(k)}, Z_1^{(k)}, \ldots, Z_{\mathcal{N}}^{(k)})$ denotes the number of mismatches and presence/absence of cutters for all nodes. Here $\mathcal{E}$ and $\mathcal{N}$ are the number of edges and nodes, respectively. Let $(\mathbf{R_L}, \mathbf{R_R}) = ((R_L^{(1)}, R_R^{(1)}), (R_L^{(2)}, R_R^{(2)}), \ldots, (R_L^{(K)}, R_R^{(K)}))$ be the lengths of the left and right end regions, and $\mathbf{N_{\mathcal{A}}} = (N_{\mathcal{A}}^{(1)}, N_{\mathcal{A}}^{(2)}, \ldots, N_{\mathcal{A}}^{(K)})$ be the fragment lengths at the attached node for all loci. Under the assumption of independent loci, we get the likelihood of indel and substitution histories for the fixed tree topology $\tau$ and edge lengths $\mathbf{t_e} = (t_1, t_2, \ldots, t_{\mathcal{E}})$:

$$
\begin{aligned}
(13) \quad & p(\mathbf{h_{id}}, \mathbf{h_{sub}} \mid \tau, \mathbf{t_e}, K, \mathbf{R_L}, \mathbf{R_R}, \mathbf{N_{\mathcal{A}}}) \\
& = \prod_{k=1}^{K} p(h_{\text{id}}^{(k)}, h_{\text{sub}}^{(k)} \mid \tau, \mathbf{t_e}, R_L^{(k)}, R_R^{(k)}, N_{\mathcal{A}}^{(k)}),
\end{aligned}
$$

which is the product of likelihoods over all loci. For a particular locus $k$, the likelihood is the product of likelihoods of indel and substitution history over all edges:

$$
\begin{aligned}
(14) \quad & p(h_{\text{id}}^{(k)}, h_{\text{sub}}^{(k)} \mid \tau, \mathbf{t_e}, R_L^{(k)}, R_R^{(k)}, N_{\mathcal{A}}^{(k)}) \\
& = \prod_{e} p_e(h_e^{(k)}, M_{C(e)}^{(k)}, Z_{C(e)}^{(k)} \mid t_e, M_{P(e)}^{(k)}, Z_{P(e)}^{(k)}, R_L^{(k)}, R_R^{(k)}, N_{P(e)}^{(k)}).
\end{aligned}
$$

For each edge $e$, the likelihood $p_e(h_e^{(k)}, M_{C(e)}^{(k)}, Z_{C(e)}^{(k)} \mid t_e, M_{P(e)}^{(k)}, Z_{P(e)}^{(k)}, R_L^{(k)}, R_R^{(k)}, N_{P(e)}^{(k)})$ is calculated by formula (10) if none of the ancestral edges have killing events. Otherwise, the likelihood for this edge is 1 since all possible events could happen along it and the events are not tracked. $C(e)$ and $P(e)$ denote the child and parent node of edge $e$, respectively. Length $N_{P(e)}^{(k)}$ is determined by $N_{\mathcal{A}}^{(k)}$ and the indel history over the ancestral edges of $P(e)$.

In the process of AFLP production, if primer pairs with different additional bases are chosen in selective amplification, we get different AFLP data sets and we call a data set from a primer pair a plate. If the AFLP data are from multiple plates, under the assumptions of the Sub-ID model, we know that markers from different plates are independent. Then the likelihood of indel and substitution histories over the whole tree is the product of likelihoods like (13) over all plates.

**5. Bayesian structure.** We are interested in the posterior distribution of topologies. Let $X$ be the observed AFLP data and $Y$ denote a three dimensional matrix that could produce $X$, as described in Section 4.2. Multiple matrices $Y$ correspond to one observed data $X$. If we specify the prior for topology $\tau$ and edge lengths $\mathbf{t_e} = (t_1, t_2, \ldots, t_\mathcal{E})$, we know that $P(\tau, \mathbf{t_e}|X) \propto P(\tau, \mathbf{t_e}) P(X|\tau, \mathbf{t_e})$. But we cannot get the likelihood $P(X|\tau, \mathbf{t_e})$ analytically, which involves the integration over all possible indel and substitution histories along the tree. So we use data augmentation and consider instead the posterior $p(\tau, \mathbf{t_e}, K, \mathbf{R_L}, \mathbf{R_R}, \mathbf{N_\mathcal{A}}, \mathbf{h_{id}}, \mathbf{h_{sub}}|X)$. By Bayes' rule,

$$
\begin{aligned}
& p(\tau, \mathbf{t_e}, K, \mathbf{R_L}, \mathbf{R_R}, \mathbf{N_\mathcal{A}}, \mathbf{h_{id}}, \mathbf{h_{sub}}|X) \\
(15) \quad & \propto p(\tau, \mathbf{t_e}, K, \mathbf{R_L}, \mathbf{R_R}, \mathbf{N_\mathcal{A}}) \times p(\mathbf{h_{id}}, \mathbf{h_{sub}}|\tau, \mathbf{t_e}, K, \mathbf{R_L}, \mathbf{R_R}, \mathbf{N_\mathcal{A}}) \\
& \times p(Y|\tau, \mathbf{t_e}, K, \mathbf{R_L}, \mathbf{R_R}, \mathbf{N_\mathcal{A}}, \mathbf{h_{id}}, \mathbf{h_{sub}}) \times p(X|Y),
\end{aligned}
$$

where $p(Y|\tau, \mathbf{t_e}, K, \mathbf{R_L}, \mathbf{R_R}, \mathbf{N_\mathcal{A}}, \mathbf{h_{id}}, \mathbf{h_{sub}})$ takes value 1 or 0 depending on whether the data produced by the indel history $\mathbf{h_{id}}$ and substitution history $\mathbf{h_{sub}}$ are consistent with the assignment of AFLP values $Y$ or not, and $p(X|Y)$ takes value 1 or 0 depending on whether the assigned AFLP values $Y$ produce markers consistent with the original data set $X$ or not. So, when we specify the priors for $(\tau, \mathbf{t_e}, K, \mathbf{R_L}, \mathbf{R_R}, \mathbf{N_\mathcal{A}})$, with the description of the model and likelihood calculation in the previous section, we can use an MCMC approach to infer the posterior for topology and edge lengths.

5.1. *Prior specification.* The topology $\tau$, edge lengths $\mathbf{t_e}$ and number of loci $K$ are assumed to be independent a priori. The prior for the topology is assumed to be uniform over all possible rooted binary tree topologies and the lengths for all edges except the edge attached to the root are mutually independent exponential random variables with a common mean $\gamma$. The attached edge length is exponential with mean $\nu$. We consider two separate models for the set of loci in the ancestral genome that are capable of producing markers. The general model includes a set of ancestral loci that evolve to produce the observed data, but may contain some loci that produce no markers in the extant taxa. In contrast, the restricted model includes a restricted set of loci, each of which evolves to produce at least one marker in one extant taxon. We take a uniform distribution on a range as a prior for the number of loci $K^G$ in the general model. The effect of the range on the inference is examined in Section 7. Under the restricted model, we take a negative binomial distribution as a prior for the number of loci $K^R$. The number of markers is taken as an empirical estimate of the mean, and we take a large variance for the negative binomial distribution. We will compare the inferences from these two models in Section 7.

Since we assume that the loci are independent, given the number of loci $K$ (either $K^G$ under the general model or $K^R$ under the restricted one), the priors for $(R_L^{(1)}, R_R^{(1)}), (R_L^{(2)}, R_R^{(2)}), \ldots, (R_L^{(K)}, R_R^{(K)})$ are independent, and so are $N_\mathcal{A}^{(1)}, N_\mathcal{A}^{(2)}, \ldots, N_\mathcal{A}^{(K)}$. For each locus $k$, the lengths of the end regions $(R_L^{(k)}, R_R^{(k)})$ take values $\{(7,9), (9,7), (9,9)\}$ with expected proportions $16:16:1$. The fragment length $N$ in the intermediate region has approximately a geometric distribution with rate $\rho = 17/4^6$ under the approximation that $\{I_i^{(4)}\}$ and $\{I_i^{(6)}\}$ are independent [Luo, Hipp and Larget (2007)], where $I_i^{(4)}$ indicates the presence of sequence "TTAA" for the four bases starting at position $i$ for $i = 1, 2, \ldots, N-3$, and $I_i^{(6)}$ indicates the presence of "GAATTC" for the six bases starting at position $i$ for $i = 1, 2, \ldots, N-5$. Assume that the observable marker length is restricted within a range between $N_{\min}$ and $N_{\max}$. We take a mixture distribution with following form as a prior for $N_\mathcal{A}^{(k)}$:

$$p(N) = \begin{cases} \dfrac{1-w}{2} \dfrac{\rho(1-\rho)^{N_{\min}-N-1}}{1-(1-\rho)^{N_{\min}-1}}, & \text{if } N < N_{\min}; \\ w \dfrac{\rho(1-\rho)^{N-N_{\min}}}{1-(1-\rho)^{N_{\max}-N_{\min}+1}}, & \text{if } N \geq N_{\min} \text{ and } N \leq N_{\max}; \\ \dfrac{1-w}{2} \rho(1-\rho)^{N-N_{\max}-1}, & \text{if } N > N_{\max}. \end{cases}$$

(16)

Let $\text{Geom}(\rho)$ and $\text{TrGeom}(\rho, N)$ represent a geometric distribution with rate $\rho$ and a truncated geometric distribution truncated at $N$, respectively. A random variable following a $\text{TrGeom}(\rho, N)$ takes value $x$ ($x = 1, 2, \ldots, N$) with probability $\rho(1-\rho)^{x-1}/(1-(1-\rho)^N)$. Then the three distributions in (16) are $N_{\min} - \text{TrGeom}(\rho, N_{\min} - 1)$ on $\{1, 2, \ldots, N_{\min} - 1\}$, a geometric distribution $\text{Geom}(\rho)$ restricted on $\{N_{\min}, N_{\min}+1, \ldots, N_{\max}\}$ and a shifted geometric distribution $\text{Geom}(\rho) + N_{\max}$ on $\{N_{\max}+1, N_{\max}+2, \ldots\}$.

**6. MCMC approach.** We sample from our model posterior $p(\tau, \mathbf{t_e}, K, \mathbf{R_L}, \mathbf{R_R}, \mathbf{N_\mathcal{A}}, \mathbf{h_{id}}, \mathbf{h_{sub}} | X)$ using reversible jump Markov chain Monte Carlo (RJMCMC) [Green (1995, 2003)]. The states that we need to update include the number of loci $K$, the assignment of AFLP values $Y$, the lengths of the end regions $(\mathbf{R_L}, \mathbf{R_R})$, the tree topology $\tau$, edge lengths $\mathbf{t_e}$, the indel history $\mathbf{h_{id}}$ and substitution history $\mathbf{h_{sub}}$. The updates involving indel histories and the number of loci change the dimension of state space, and we follow Green (1995) to propose reversible jump updates and calculate the acceptance probabilities. Our MCMC algorithm employs a deterministic-scan line Metropolis-within-Gibbs [Tierney (1994)] approach. Fixing the topology and edge lengths, we update the number of loci by adding a new locus (including new indel and substitution histories) or deleting an existing locus (including the old indel and substitution histories for this locus). Fixing the number of loci $K$, we update the assignment of AFLP values $Y$, the lengths of the end regions $(R_L^{(k)}, R_R^{(k)})$ and fragment length $N_{\mathcal{A}}^{(k)}$ for each locus $k = 1, 2, \ldots, K$. Fixing $K$, $(\mathbf{R_L}, \mathbf{R_R})$ and $Y$, we update the indel history $h_{\text{id}}^{(k)}$ and substitution history $h_{\text{sub}}^{(k)}$ for each locus $k = 1, 2, \ldots, K$. Except for the substitution status, all other updates involve proposing new indel histories compatible with the AFLP data. We will focus on the update of indel histories on a single edge, given whether or not the history contains a killing event. This is the essential part of most updates involving indel histories. Luo (2007) describes all the updates in detail.

6.1. *Update an indel history without killing events.* We denote the edge length as $T$. Updating an indel history without killing events, we need to propose a new history containing no killing events with fragment length $N_0$ and $N_T$ at the parent and child node, respectively. The idea is to first sample a potential time for the next indel event from an exponential distribution with a rate that accounts for the rates of indel events in the intermediate region. If the cumulative time is less than the edge length $T$, we propose an insertion or deletion, according to their rates in the intermediate region, that does not destroy the end regions; otherwise, we check whether the fragment length matches $N_T$ or not. If the new fragment length after these proposed indel events is not $N_T$, we propose an additional event to match the length. The detailed proposal is described in the Appendix.

If we let $k$ be the number of indel events proposed, and let $h = (t_i, w_i, s_i, l_i, i = 1, \ldots, k)$, then the proposal density under this scheme is

$$
\begin{aligned}
q(h) = {} & \exp\left(-\sum_{i=1}^{k-1} \zeta_i(t_i - t_{i-1})\right) \\
& \times \prod_{i=1}^{k-1}\left\{\left(I(w_i > 0)\lambda + \frac{\mu I(w_i < 0)}{1 - (1-r)^{R_L + N_{i-1} - s_i}}\right)g(l_i)\right\} \\
& \times \exp(-\zeta_k(t_{k+1} - t_{k-1})) \\
(17)\quad & \times \frac{1}{t_{k+1} - t_{k-1}} \times \left(\frac{I(w_k > 0)}{N_{k-1} + 1} + \frac{I(w_k < 0)}{N_k + 1}\right) \\
& + \exp\left(-\sum_{i=1}^{k+1} \zeta_i(t_i - t_{i-1})\right)
\end{aligned}
$$

$$\times \prod_{i=1}^{k}\left\{\left(I(w_i>0)\lambda + \frac{\mu I(w_i<0)}{1-(1-r)^{R_L+N_{i-1}-s_i}}\right)g(l_i)\right\},$$

where $\zeta_i = (N_{i-1}+1)\lambda + N_{i-1}\mu$, and $g(\cdot)$ denotes the probability mass of a geometric distribution with mean $1/r$. The first term corresponds to the case that the last event is specifically proposed to match $N_T$. The second one corresponds to the case when the proposed $k$ events happen to match $N_T$. Letting the new and old indel histories be $h'$ and $h$, respectively, the Jacobian of transformation is 1 [Green (1995)], and the acceptance probability is

$$\text{(18)} \qquad \min\left\{1, \frac{p(h'|N_0)}{p(h|N_0)} \frac{q(h)}{q(h')}\right\},$$

where the likelihoods $p(h'|N_0)$ and $p(h|N_0)$ can be calculated according to formula (8), and the proposal density is given in (17).

6.2. *Update an indel history containing a killing event.* To propose an indel history containing a killing event, we repeatedly propose indel events along the edge according to the likelihood. If a killing event is proposed, then we stop. If no killing event is proposed when the cumulative time exceeds the edge length, we propose to add one killing event after the last event where the cumulative time is less than the edge length. The added killing event takes any of the four types $-1_R$, $-1_P$, $-1_N$ and $1_R$ as described in Table 2 with probabilities proportional to their rates. Suppose that $k$ events $h = (t_i, w_i, s_i, l_i, i=1,\ldots,k)$ are proposed. Then the proposal density is

$$\text{(19)} \quad \begin{aligned} q(h) &= \exp\left(-\sum_{i=1}^{k} \eta_i(t_i - t_{i-1})\right) \prod_{i=1}^{k} \{(I(w_i>0)\lambda + I(w_i<0)\mu)g(l_i)\} \\ &\quad + \exp\left(-\sum_{i=1}^{k-1} \eta_i(t_i - t_{i-1})\right) \times \frac{\exp(-\eta_k(t_{k+1}-t_{k-1}))}{t_{k+1}-t_{k-1}} \\ &\quad \times \prod_{i=1}^{k} \{(I(w_i>0)\lambda + I(w_i<0)\mu)g(l_i)\} \\ &\quad \times (1-r)^{(-R_L - N_{k-1} + s_{k-1}) \times I(w_k = -1_N)}. \end{aligned}$$

Letting the new and old indel histories be $h'$ and $h$, respectively, the acceptance probability is given by (18), where the likelihoods $p(h'|N_0)$ and $p(h|N_0)$ can be calculated according to formula (9), and the proposal densities are given in (19).

**7. Simulation study.** We study the effects of different prior settings on our Bayesian inference with several simulated data sets of distinct number of taxa or loci. We illustrate the results from one simulated data. The results are consistent with other simulation studies. Table 4 contains one simulated data set from topology $((A, B), ((C, D), (E, F)))$ and ten loci, which leads to the observed AFLP data given in Table 5. To see the effects of the priors for the number of loci on the inference, we take three priors: $K^G \sim \text{Unif}\{1, 2, \ldots, 15\}$, $K^G \sim \text{Unif}\{1, 2, \ldots, 50\}$ and

$$(20) \qquad K^R \sim \text{NegBinom}(\mu_K, 1000),$$

where $\mu_K$ is taken as the number of markers for the data set. To study the effect of priors for fragment length at the attached node $N_\mathcal{A}$, we take two priors: (16) with $w = 0.95$ and prior

$$(21) \quad p(N) = \begin{cases} \dfrac{1-w}{2} \dfrac{\rho(1-\rho)^{N_{\min}-N-1}}{1-(1-\rho)^{N_{\min}-1}}, & \text{if } N < N_{\min}; \\ \dfrac{w}{N_{\max} - N_{\min} + 1}, & \text{if } N \geq N_{\min} \text{ and } N \leq N_{\max}; \\ \dfrac{1-w}{2} \rho(1-\rho)^{N-N_{\max}-1}, & \text{if } N > N_{\max}; \end{cases}$$

with $w = 0.9$.

TABLE 4
*An example of simulated data Y from 10 loci. For any locus k and any taxon i, marker information is shown with two values, where the first ($y_{ki1} = 1$, 0 or $-1$) indicates whether or not the ith taxon produces a marker with length $y_{ki2}$ (the second value). Loci 1, 5 and 6 do not produce markers. Loci 7, 8 and 9 each produce two markers (locus-splitting)*

|   | **Locus 1** | | **Locus 2** | | **Locus 3** | | **Locus 4** | | **Locus 5** | |
|---|---|---|---|---|---|---|---|---|---|---|
| A | 0 | 54 | 1 | 61 | 1 | 76 | 1 | 111 | 0 | 122 |
| B | 0 | 54 | 1 | 61 | 1 | 76 | 1 | 111 | 0 | 122 |
| C | 0 | 54 | 1 | 61 | 1 | 76 | 1 | 111 | 0 | 122 |
| D | 0 | 54 | 1 | 61 | 1 | 76 | 1 | 111 | 0 | 122 |
| E | 0 | 54 | 1 | 61 | 0 | 76 | 0 | 111 | 0 | 122 |
| F | 0 | 54 | 1 | 61 | 0 | 76 | 0 | 111 | 0 | 122 |
|   | **Locus 6** | | **Locus 7** | | **Locus 8** | | **Locus 9** | | **Locus 10** | |
| A | 0 | 127 | 0 | 135 | 1 | 216 | 1 | 219 | 1 | 412 |
| B | 0 | 127 | 0 | 135 | 1 | 216 | 1 | 219 | 1 | 412 |
| C | 0 | 127 | 1 | 136 | 1 | 215 | 1 | 221 | 0 | 412 |
| D | 0 | 127 | 1 | 136 | 1 | 215 | 1 | 221 | 0 | 412 |
| E | $-1$ | — | 1 | 137 | 1 | 215 | 0 | 221 | 0 | 410 |
| F | $-1$ | — | 1 | 137 | 1 | 215 | 0 | 221 | 0 | 410 |

TABLE 5
*The observable AFLP data X from the simulated data in Table 4*

| Marker length | 61 | 76 | 111 | 136 | 137 | 215 | 216 | 219 | 221 | 412 |
|---|---|---|---|---|---|---|---|---|---|---|
| A | 1 | 1 | 1 | 0 | 0 | 0 | 1 | 1 | 0 | 1 |
| B | 1 | 1 | 1 | 0 | 0 | 0 | 1 | 1 | 0 | 1 |
| C | 1 | 1 | 1 | 1 | 0 | 1 | 0 | 0 | 1 | 0 |
| D | 1 | 1 | 1 | 1 | 0 | 1 | 0 | 0 | 1 | 0 |
| E | 1 | 0 | 0 | 0 | 1 | 1 | 0 | 0 | 0 | 0 |
| F | 1 | 0 | 0 | 0 | 1 | 1 | 0 | 0 | 0 | 0 |

Table 6 summarizes the inferences for topologies and the number of loci under different prior settings. In the six simulations under the Sub-ID model, the true topology and the true number of loci producing the AFLP data have the highest posterior probabilities. For comparison, the substitution-only model supports $(((A,B),(C,D)),(E,F))$ with higher posterior probability than the true topology, and infers clade $((A,B),(C,D))$ with higher frequency (0.501) than $((C,D),(E,F))$ (0.333). In an examination of the effects of priors (16) and (21) for the fragment size at the ancestral node $N_\mathcal{A}$, we find that this choice had no detectable effect on the posterior distribu-

TABLE 6
*Posterior inferences (in %) for topologies and $K^R$ under different prior settings. Values in parentheses are standard errors (in %). True topology and $K^R$ are shown in bold face. $N_{\mathcal{A},G}$ and $N_{\mathcal{A},U}$ represent the priors for $N_\mathcal{A}$ as given by (16) and (21), respectively*

|  | $K^G \sim$ Unif$\{1,2,\ldots,15\}$ | | $K^G \sim$ Unif$\{1,2,\ldots,50\}$ | | $K^R \sim$ NB(10,1000) | | Sub- |
|---|---|---|---|---|---|---|---|
|  | $N_{\mathcal{A},U}$ | $N_{\mathcal{A},G}$ | $N_{\mathcal{A},U}$ | $N_{\mathcal{A},G}$ | $N_{\mathcal{A},U}$ | $N_{\mathcal{A},G}$ | model |
| Topology |  |  |  |  |  |  |  |
| **((A,B),((C,D),(E,F)))** | 32.4 (0.3) | 31.5 (0.4) | 28.7 (0.1) | 31.0 (1.1) | 33.8 (0.2) | 32.7 (0.5) | 16.3 (0.1) |
| ((A,B),((C,(E,F)),D)) | 20.1 (0.3) | 19.3 (0.2) | 20.4 (0.0) | 19.9 (0.3) | 20.3 (0.1) | 21.1 (0.2) | 4.8 (0.1) |
| ((A,B),(C,(D,(E,F)))) | 20.8 (0.5) | 19.5 (0.5) | 20.4 (0.4) | 19.4 (0.5) | 20.6 (0.4) | 21.4 (0.2) | 5.0 (0.1) |
| (((A,B),(E,F)),(C,D)) | 8.7 (0.6) | 10.5 (0.4) | 11.0 (0.9) | 11.7 (1.5) | 10.0 (0.7) | 9.9 (0.5) | 6.9 (0.1) |
| (((A,B),(C,D)),(E,F)) | 3.7 (0.4) | 4.3 (0.3) | 5.9 (0.3) | 4.9 (0.2) | 4.4 (0.3) | 4.4 (0.1) | 21.3 (0.2) |
| ((A,(B,(E,F))),(C,D)) | 0.7 (0.0) | 0.8 (0.1) | 1.1 (0.1) | 0.8 (0.1) | 0.3 (0.1) | 0.4 (0.0) | 0.1 (0.0) |
| (((A,(E,F)),B),(C,D)) | 0.7 (0.1) | 0.7 (0.1) | 1.0 (0.1) | 1.0 (0.0) | 0.5 (0.1) | 0.4 (0.0) | 0.2 (0.0) |
| ((A,((C,D),(E,F))),B) | 1.4 (0.1) | 1.3 (0.1) | 0.9 (0.0) | 0.8 (0.3) | 0.7 (0.0) | 0.7 (0.1) | 2.3 (0.0) |
| (A,(B,((C,D),(E,F)))) | 1.4 (0.1) | 1.3 (0.1) | 0.9 (0.1) | 1.0 (0.4) | 0.8 (0.1) | 0.7 (0.1) | 2.3 (0.0) |
| Cumulative prob. |  |  |  |  |  |  |  |
| (%) | 90.7 | 89.4 | 90.2 | 90.6 | 92.3 | 92.4 | 59.2 |
| $K^R$ |  |  |  |  |  |  |  |
| **7** | 93.7 (0.3) | 87.8 (0.1) | 89.2 (0.5) | 84.1 (0.3) | 91.0 (0.3) | 85.4 (0.4) | — |
| 8 | 6.1 (0.3) | 11.4 (0.1) | 10.1 (0.4) | 14.4 (0.2) | 8.7 (0.2) | 13.4 (0.4) | — |
| 9 | 0.2 (0.0) | 0.7 (0.0) | 0.7 (0.1) | 1.3 (0.1) | 0.3 (0.1) | 1.2 (0.1) | — |

Table 7

*Proportions (in %) of pairs of markers produced by a single locus in MCMC samples with different priors for the number of loci and $N_\mathcal{A}$. Pairs of markers in italized face are produced by single loci. Values in parentheses are the estimated Monte Carlo standard errors for proportions (in %)*

|  | *(136, 137)* | *(215, 216)* | *(219, 221)* | (215, 219) | (216, 221) |
|---|---|---|---|---|---|
| Unif$\{1, 2, \ldots, 15\}$, $N_{\mathcal{A},U}$ | 99.63 (0.1) | 98.30 (1.1) | 96.18 (1.1) | 1.20 (1.1) | 1.04 (1.0) |
| Unif$\{1, 2, \ldots, 15\}$, $N_{\mathcal{A},G}$ | 99.22 (0.0) | 97.42 (1.8) | 94.45 (1.0) | 1.62 (1.5) | 1.53 (1.5) |
| Unif$\{1, 2, \ldots, 50\}$, $N_{\mathcal{A},U}$ | 98.92 (0.1) | 96.61 (1.1) | 92.77 (1.1) | 2.47 (1.0) | 2.11 (1.0) |
| Unif$\{1, 2, \ldots, 50\}$, $N_{\mathcal{A},G}$ | 98.69 (0.3) | 97.67 (1.0) | 94.08 (1.1) | 1.10 (0.8) | 0.86 (0.8) |
| NB(10, 1000), $N_{\mathcal{A},U}$ | 99.52 (0.1) | 96.31 (0.0) | 93.35 (0.3) | 2.79 (0.2) | 2.37 (0.1) |
| NB(10, 1000), $N_{\mathcal{A},G}$ | 99.35 (0.1) | 96.67 (0.8) | 93.19 (0.7) | 2.27 (0.8) | 1.93 (0.8) |

tion of the tree topology, but that the posterior distribution of the inferred number of loci $K^R$, that produce markers under the restricted model, is relatively sensitive to this choice of prior distribution. Changing the prior for the number of loci does not affect the inference for topologies and $K^R$ greatly.

The Sub-ID model also allows us to infer which pairs of markers are possibly produced by single loci. We know that markers from each of the three pairs (136, 137), (215, 216) and (219, 221) are homologous. From the MCMC samples, we obtain the proportions that each pair is produced by a single locus and list them in Table 7. All six simulations infer that each of the three pairs of markers (136, 137), (215, 216) and (219, 221) is produced by a single locus with probabilities over 92% and that each of the two pairs of markers (215, 219) and (216, 221) is much less likely homologous (with probabilities less than 3%). We know that these pairs are actually produced by different loci.

7.1. *Sensitivity of the Jukes–Cantor assumption.* We assume the Jukes–Cantor model for nucleotide substitution, which allows us to analytically calculate the transition probabilities for the number of mismatches ($M$) and the presence/absence of cutters ($Z$). This assumption is the simplest model for nucleotide substitution and the real evolutionary process may have different base frequencies and transition/transversion rates. We can study the robustness of our Sub-ID model by applying our phylogenetic inference approach to data simulated from insertion/deletion and nonJC model for nucleotide substitution. We first consider the Tamura–Nei (TN) model [Tamura and Nei (1993)] with rate variations among sites. The TN model assumes different base frequencies and allows different rates for transversion, transition between purines and transition between pyrimidines. The substitution rates at different sites are assumed to be different and independently follow a gamma distribution (TN + $\Gamma$ model) [Yang (1993, 1994)]. Here, we assume the rates to follow Gamma(0.2, 0.2) and specify the base frequencies according to the nucleotide proportions in the genome of rice. Denote the model with the insertion/deletion process as described in this paper and with the $TN + \Gamma$ model for substitution process as $TN + \Gamma + ID$ model. We apply our methodology to the AFLP data simulated from the $TN + \Gamma + ID$ model to check the sensitivity of the Jukes–Cantor assumption. We take (16) as the prior for the fragment length at the attached node $\mathcal{A}$ and (20) or $K^G \sim \text{Unif}\{1, 2, \ldots, 1000\}$ as priors for the number of loci. The posterior probabilities of topologies from the Sub-ID model with different priors for the number of loci are close (Table 8), with the negative binomial prior (20) giving a little stronger support for the true topology (0.541) than the uniform prior (0.520). For comparison, we list the posterior probabilities inferred from the substitution-only model in Table 8, which are more diverse than the inferences from the Sub-ID model. Posterior inference for the number of loci $K^R$ are close from the two simulations (figures not shown). Applying the uniform prior, we inferred the true values of $K^R$ (20 and 23 for two plates) with the highest posterior probability. Using the negative binomial prior (20), we inferred 24 with a little higher probability (0.363) than the true value (with probability 0.353) for the second plate.

TABLE 8
*Posterior probabilities (in %) for topologies analyzing data simulated from $TN+\Gamma+ID$ model. Values in parentheses are Monte Carlo standard errors (in %) estimated from multiple independent MCMC runs*

|  | Sub-ID model | | |
|---|---|---|---|
|  | $K^G \sim \text{Unif}$ | $K^R \sim \text{NB}$ | Sub-model |
| **((A,B),((C,D),(E,F)))** | 52.0 (4.7) | 54.1 (4.0) | 21.8 (0.1) |
| (((A,B),(C,D)),(E,F)) | 14.7 (0.7) | 17.5 (0.1) | 21.3 (0.3) |
| (((A,B),(E,F)),(C,D)) | 9.1 (2.4) | 13.7 (1.6) | 15.2 (0.4) |
| ((A,((C,D),(E,F))),B) | 13.4 (1.2) | 10.3 (1.5) | 17.1 (0.1) |
| (A,(B,((C,D),(E,F)))) | 5.9 (0.9) | 3.3 (0.9) | 3.5 (0.0) |
| Cumulative prob. (%) | 95.1 | 98.9 | 78.9 |

The previous results are consistent with results from the analysis of another data set. Both studies inferred the true topology and the number of loci $K^R$ producing markers with high probabilities, implying that our method seems to be robust to likelihood model misspecification about the substitution process.

**8. Case study.** We applied our methodology to a data set with fourteen different sedge species and two plates that include 62 and 64 markers, respectively. This is a subset of a larger data set published in Hipp et al. (2006). The taxa with number of individuals from each are as follows: *Carex bebbii* (1), *C. bicknellii* (1), *C. festucacea* (2), *C. normalis* (2), *C. oronensis* (2), *C. tenera* var. *echinodes* (2), *C. tenera* var. *tenera* (2), and *C. tincta* (2). The taxa chosen for this study represent a morphologically cohesive clade, with two closely-related taxa as outgroup (*C. bebbii* and *C. bicknellii*). Monophyly of the former is supported by neighbor joining (NJ) and minimum evolution (ME) analyses on an expanded data set that includes all members of an eastern North American clade identified in a previous study using nuclear ribosomal DNA sequence data. Some of the relationships within the group, however, are not strongly supported using distance methods, which was one of the interests in exploring the phylogeny of this group using a more realistic model of character evolution.

We ran four Metroplis-Coupled MCMC procedures, each with three heated chains and one cold chain, from different starting points. Each run had 1,000,000 iterations and we sampled every 100 iterations from the cold chain. It took about 130 hours to run a MCMCMC procedure. Most of the acceptance ratios are in the range of $[0.1, 0.4]$, except that the updates of root position and the lengths of the end regions $(R_L, R_R)$ were accepted less frequently (around 0.08 or 0.09), and the updates of AFLP specification $(Y)$ and indel histories without killing event are more easily accepted (about 0.8 or 0.9). To assess convergence for continuous parameters, we computed Gelman–Rubin R statistics [Gelman and Rubin (1992)] for sampled leaf edge lengths and indel parameters. Internal edge lengths cannot be used since they do not necessarily retain definition across topologies. All statistics are near 1 with deviation from 1 less than 0.05.

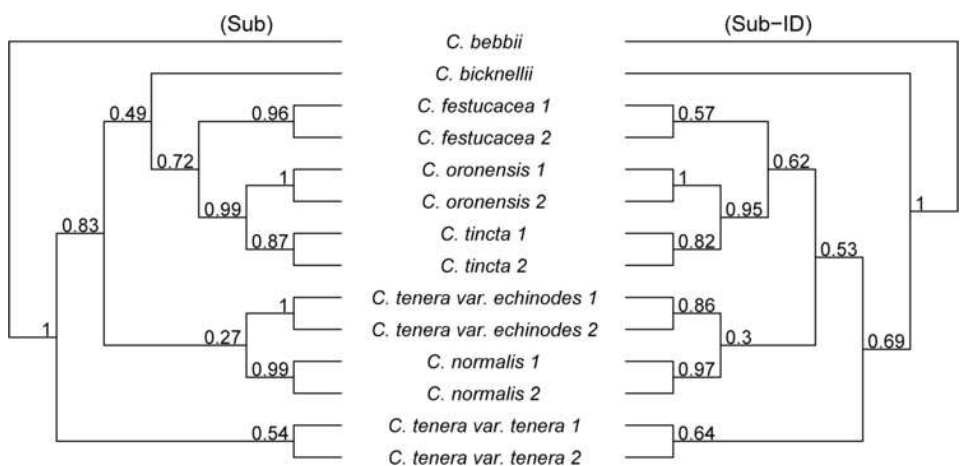

Fig. 2. *Priority consensus trees from Sub-ID model (right) and substitution-only model (left). Numbers represent posterior probabilities of each clade. The trees are rooted such that C. Bebbii is an outgroup.*

We summarize the sampled topologies with the priority consensus tree (Figure 2), which is a fully resolved tree built up by sorting groups in descending order by posterior probabilities, and including clades under the restriction that clades with lower probabilities do not contradict with those with higher probabilities. For comparison, we include the consensus tree inferred with the substitution-only model in Figure 2. In both simulations, individuals from the same species were grouped together with probability over 54%, and both consensus trees contain clade {*C. festucacea*, *C. oronensis*, *C. tincta*} with probability over 62%. We will call this clade F/O/Ti. Both trees contain clade {*C. tenera* var. *echinodes*, *C. normalis*} (denoted as Te/N) with small probabilities (27% for substitution-only model and 30% for Sub-ID model). The two topologies differ in the grouping of taxon *C. bicknellii* (denoted as Bi) and clade *C. tenera* var. *tenera* (denoted as Tt). The substitution-only model first group Bi with clade F/O/Ti with probability 49%, and take Tt as a sister group with clade Bi/F/O/Ti/Te/N, while the Sub-ID model infers Bi as an outgroup of clade F/O/Ti/Te/N/Tt, which is consistent with the inferences from Luo, Hipp and Larget (2007) using the NJ method, MrBayes method or substitution-only model on a larger data set with 9 plates that take Bi as an outgroup of clade F/O/Ti/Te/N/Tt. The consensus tree from the Sub-ID model inferred from the smaller data set with two plates has the same topology as the most probable tree (which is also the priority consensus tree) from the substitution-only model (with probability 68.8%) applied on a larger data set with nine plates [see Figure 6 in Luo, Hipp and Larget (2007)].

The priority consensus tree from the Sub-ID model in Figure 2 is the most probable tree [Figure 3(c)] and has posterior probability 6.4%. It has small support for clade Te/N (with posterior probability 30%), and clade Te/N is split in the second most probable tree (with posterior probability 4.7%). The second most probable trees from both inferences [Figure 3(b,d)] have the same topology. For the substitution-only model, the most probable tree [Figure 3(a)] differs from the consensus tree by splitting clade Te/N.

The Sub-ID model allows us to infer which set of markers are possibly produced by a single locus. Table 9 lists several pairs of markers for which our analysis indicates evidence of potential locus-splitting. In addition, we get the probability for each marker that it is a superposition. The probability ranges from 1.25% to 67.72%. On average, 17.23% of the markers are superpositions.

We can also infer the insertion and deletion rates relative to the substitution rate. Taking the substitution rate as 1 and applying priors $r \sim \text{Unif}(0, 1)$, $\mu \sim \text{Gamma}(4, 100)$ and $\beta \sim \text{Beta}(3, 1)$ independently, where $\lambda = \mu\beta \sim \text{Gamma}(3, 100)$, we infer that the posterior means for insertion and deletion rates are 0.025 and 0.031, with 95% credible intervals $(0.013, 0.041)$ and $(0.017, 0.049)$, respectively. The 95% credible interval for $r$ is $(0.044, 0.316)$.

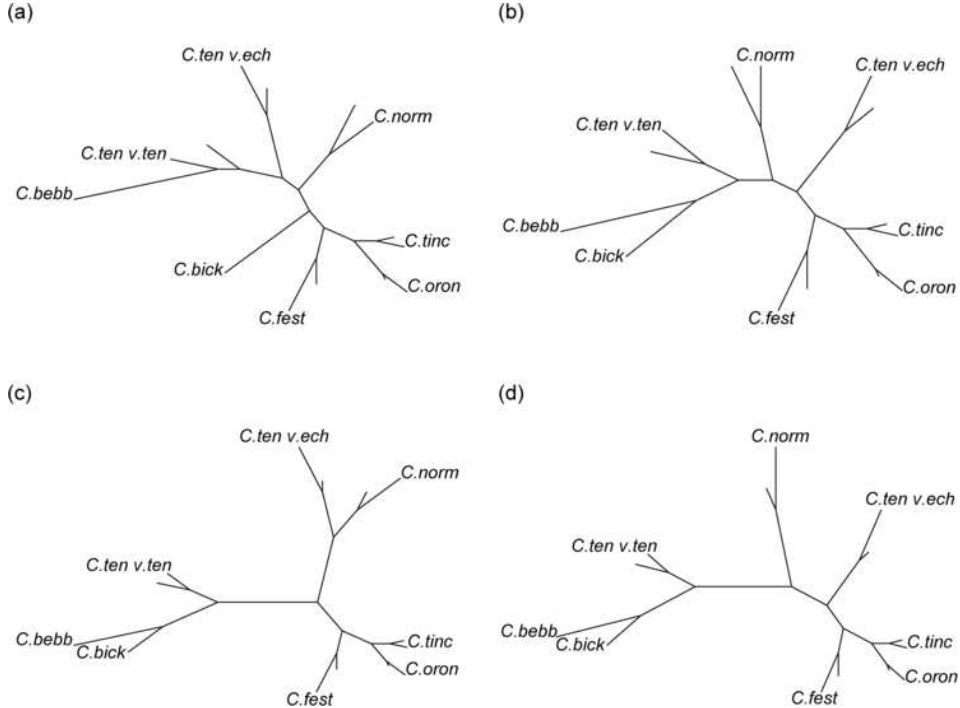

FIG. 3. *The first two most probable trees from the substitution-only model* (a,b) *and the Sub-ID model* (c,d).

While the substitution rate per site is much higher than the indel rate per site, both substitution and indel processes are important for AFLP marker evolution. From equations (3) and (7), first order approximations of the rates of events causing marker loss or change are $1 \times (R + N/64)$ for substitutions and $(\lambda + \mu) \times (R + N)$ for indels. In this data set with an estimated combined indel rate of 0.056, indels are expected to account for about 10% of changes in short fragments and more than half the changes in the longest fragment. The rate estimations may be sensitive to prior specification, and we have not examined this possibility in detail.

**9. Discussion.** We have described a model for AFLP evolution involving both nucleotide substitution and insertion/deletion, and have developed a Bayesian approach to infer phylogenies. Compared to the substitution-only model, modeling the indel process relaxes the assumption that markers are independent and homologous. In addition to inferring the topology, we can infer the subset of markers that are produced by a single locus in the MCMC samples and which set of markers are possibly homologous. Furthermore, our method provides an estimate of the genome-wide indel rate relative to the substitution rate from AFLP marker data. The phylogenetic inference based on the Sub-ID model takes more time and can have smaller support for some clades than the Bayesian method based on the substitution-only model of Luo, Hipp and Larget (2007). Since the Sub-ID model is more complicated, the likelihood calculation is more time expensive and more MCMC updates are needed. However, when we applied both methods to a subset of a larger data set, and compared the inferred consensus trees with that obtained by applying the substitution-only model on a larger data set [the tree obtained from Luo, Hipp and Larget (2007)], the Sub-ID model inferred the same consensus tree as the substitution-only model on the larger data set, while the substitution-only model infers different consensus trees on the smaller and larger data sets. The Sub-ID model better captures the phylogenetic information contained in AFLP marker data.

TABLE 9
*Marker pairs produced by a single locus with high proportions in the MCMC samples. "Prob. (%)" represents the percentage of the MCMC samples in which the marker pairs are produced by a single locus, and "se (%)" shows the standard errors of the percentages*

|  | Plate 1 | | | | | | | | | | | | Plate 2 | | | | | | | |
|---|---|---|---|---|---|---|---|---|---|---|---|---|---|---|---|---|---|---|---|---|
| Markers | 476 | 477 | 364 | 365 | 236 | 237 | 297 | 298 | 111 | 113 | 237 | 251 | 94 | 100 | 111 | 112 | 177 | 179 | 130 | 131 |
| C. bebb | 0 | 0 | 0 | 0 | 0 | 0 | 1 | 0 | 0 | 1 | 0 | 0 | 0 | 0 | 1 | 0 | 0 | 1 | 0 | 1 |
| C. bick | 1 | 0 | 1 | 0 | 0 | 0 | 0 | 1 | 0 | 0 | 0 | 0 | 1 | 0 | 1 | 0 | 0 | 1 | 0 | 1 |
| C. fest1 | 0 | 0 | 1 | 0 | 0 | 0 | 0 | 1 | 1 | 0 | 0 | 0 | 0 | 1 | 1 | 0 | 0 | 1 | 0 | 0 |
| C. fest2 | 0 | 1 | 1 | 0 | 0 | 0 | 0 | 1 | 1 | 0 | 0 | 0 | 1 | 0 | 1 | 0 | 0 | 1 | 1 | 0 |
| C. oron1 | 0 | 0 | 1 | 0 | 1 | 0 | 1 | 0 | 1 | 0 | 0 | 1 | 1 | 0 | 1 | 0 | 0 | 1 | 0 | 1 |
| C. oron2 | 0 | 0 | 1 | 0 | 0 | 0 | 1 | 0 | 1 | 0 | 0 | 1 | 1 | 0 | 1 | 0 | 0 | 1 | 0 | 1 |
| C. tinc1 | 1 | 0 | 0 | 1 | 0 | 0 | 0 | 0 | 1 | 0 | 0 | 0 | 1 | 0 | 1 | 0 | 1 | 0 | 0 | 1 |
| C. tinc2 | 0 | 0 | 0 | 1 | 0 | 0 | 0 | 0 | 1 | 0 | 0 | 0 | 1 | 0 | 1 | 0 | 1 | 0 | 0 | 0 |
| C. norm1 | 1 | 0 | 1 | 0 | 0 | 0 | 0 | 0 | 1 | 0 | 0 | 0 | 0 | 0 | 0 | 1 | 0 | 1 | 0 | 0 |
| C. norm2 | 1 | 0 | 1 | 0 | 0 | 0 | 0 | 0 | 1 | 0 | 0 | 0 | 0 | 0 | 0 | 1 | 0 | 1 | 0 | 0 |
| C. echi1 | 1 | 0 | 1 | 0 | 0 | 1 | 1 | 0 | 1 | 0 | 1 | 0 | 0 | 0 | 1 | 0 | 0 | 1 | 0 | 0 |
| C. echi2 | 1 | 0 | 1 | 0 | 0 | 0 | 1 | 0 | 1 | 0 | 0 | 0 | 0 | 0 | 0 | 0 | 0 | 1 | 0 | 1 |
| C. tene1 | 1 | 0 | 0 | 1 | 0 | 0 | 0 | 0 | 1 | 0 | 0 | 0 | 1 | 0 | 1 | 0 | 0 | 1 | 0 | 0 |
| C. tene2 | 1 | 0 | 0 | 1 | 0 | 0 | 0 | 0 | 1 | 0 | 0 | 0 | 1 | 0 | 1 | 0 | 0 | 1 | 0 | 0 |
| Prob. (%) | 77.03 | | 76.47 | | 26.97 | | 24.98 | | 21.27 | | 16.96 | | 64.74 | | 63.03 | | 43.71 | | 27.14 | |
| se (%) | 3.55 | | 4.76 | | 7.47 | | 0.92 | | 3.36 | | 3.46 | | 1.45 | | 5.13 | | 6.14 | | 2.13 | |

We considered a general and a restricted model for the number of loci. Different priors on the number of loci $K^G$ in the general set or $K^R$ in the restricted set do not affect the inference of topologies greatly, but have more effect on the posterior inference of the number of loci producing markers. Simulation study shows that our method recovers the true topology and true number of loci producing markers with high probability. Using a uniform prior of $K^G$, we can further infer the posterior distribution for the number of loci at node $\mathcal{A}$. The inferences of $K^G$ are indistinguishable when different uniform distributions are applied with ranges large enough (figure not shown). Using a negative binomial distribution as a prior for $K^R$, MCMC simulations mix faster than those with a uniform prior for $K^G$. Different priors on the fragment length $N_\mathcal{A}$ at node $\mathcal{A}$ have more effect on the inference of $K^R$ than on topology.

Assumption of the Jukes–Cantor model for nucleotide substitution simplifies the model in that the mismatch processes $M(t)$ for the end regions is Markovian and the process $Z(t)$ for the intermediate region is approximately a Markov process. Hence, we do not need to record all nucleotide bases in a fragment with length $N$, but just keep the number of mismatches in the end regions and the presence/absence of cutters in the intermediate region. Thus, if we consider the substitution process only, we reduce the number of states from $N$ to 2, and reduce the number of all possible values for the states from $4^N$ to $R+3$, where $R$ is the length of end regions. The real substitution process may not have equal base frequencies and transition rates, which are what the Jukes–Cantor model assumes. The sensitivity study by inferring topologies from data simulated from a nonJC model (like $TN + \Gamma + ID$) reveals that our Bayesian methodology is robust and can recover the true topology and the number of loci producing markers $K^R$.

When an end region is destroyed by a killing event in the SubID model, it is destroyed forever and we have not allowed the rebirth of killed fragments. Hence, the indel process is nonreversible and a rooted tree is required to model the indel history. To know whether or not a new indel or substitution event remedies an end region and changes a killing to a nonkilling event, we need to know the bases in the destroyed end region and its neighboring bases, but this information is unobtainable in the current Sub-ID model since we do not record the full sequences. Additional effort is needed to build up a model allowing birth of new AFLP markers.

The expensive computational cost of our method limits the scope of the data set to which our method may be applied. The Bayesian approach described in Luo, Hipp and Larget (2007) requires large computation [calculations of $2 \times 2$, $34 \times 34$ and $38 \times 38$ matrices for processes $Z(t)$ and $M(t)$ on each branch] for likelihood calculations, and a full computation for each marker must be carried out since marker length is used in the model and computations among markers with identical patterns cannot be shared. In addition to this, the Bayesian method based on the Sub-ID model faces more computation problems. We can calculate the likelihood of a topology under the substitution-only model using the pruning algorithm [Felsenstein (2004)], but we cannot do this under the Sub-ID model since the likelihood involves the integration over all possible substitution and indel histories. We use the data augmentation technique and MCMC method to avoid the direct calculation of likelihood of a topology, but they raise new questions of proposing good indel histories and changing topologies. When a topology is proposed to change, we must propose new indel and substitution histories to fit the new topology. In our implementation, to increase the acceptance rate, we propose a new topology with a local update and propose new indel histories by keeping as much of the old histories as possible so that the new and old histories have closer likelihoods and proposal densities and the acceptance probability is closer to 1. Broad application of our Bayesian approach based on the Sub-ID model in phylogenetic inference relies on better MCMC update algorithms and the improvement of software implementation.

It may be important to know the relative phylogenetic information contained in AFLP data as opposed to aligned sequence data. This is a complicated issue, the solution of which will depend on factors such as the overall rate of substitutions, substitution rate variation among different genes, genome size, the relative rate of indels to substitutions, the nature of the indel process, and specifics of the underlying tree among other considerations. AFLP marker data is relatively inexpensive, but it remains unclear, in general, how the larger AFLP data set would compare to a smaller DNA sequence available at equal cost. We speculate that AFLP markers which measure changes on a genomic scale may be most advantageous relative to aligned DNA sequences in situations where the rate of substitution in standardly available single gene sequences is so low that the aligned sequences contain very little information.

## APPENDIX: PROPOSE AN INDEL HISTORY WITHOUT KILLING EVENTS

Set $t_0 = 0$, $i = 0$;
Repeat:
    set $\zeta_{i+1} = (N_i + 1)\lambda + N_i\mu$, $\Delta t \sim \text{Exp}(\zeta_{i+1})$.
    if $t_i + \Delta t < T$, do:
        $t_{i+1} = t_i + \Delta t$;
        with probability $(N_i + 1)\lambda/\zeta_{i+1}$, propose an insertion as the $(i+1)$th event:
            $w_{i+1} = 1$;
            $s_{i+1} \sim \text{Unif}\{R_L, R_L + 1, \ldots, R_L + N_i\}$;
            $l_{i+1} \sim \text{Geom}(r)$;
        with probability $N_i\mu/\zeta_{i+1}$, propose a deletion as the $(i+1)$th event:
            $w_{i+1} = -1$;
            $s_{i+1} \sim \text{Unif}\{R_L, R_L + 1, \ldots, R_L + N_i - 1\}$;
            $l_{i+1} \sim \text{TrGeom}(r, N_i + R_L - s_{i+1})$;
        $N_{i+1} = N_i + w_{i+1} \times l_{i+1}$;
        set $i \leftarrow i + 1$.
    else, do:
        if $N_i = N_T$, Stop.

else, do:
$t_{i+1} \sim \text{Unif}(t_i, T)$;
if $N_i < N_T$, set the last event as an insertion:
$w_{i+1} = 1$, $l_{i+1} = N_T - N_i$, $s_{i+1} \sim \text{Unif}\{R_L, R_L + 1, \ldots, R_L + N_i\}$, $N_{i+1} = N_T$;
else, set the last event as a deletion:
$w_{i+1} = -1$, $l_{i+1} = N_i - N_T$, $s_{i+1} \sim \text{Unif}\{R_L, R_L + 1, \ldots, R_L + N_T\}$, $N_{i+1} = N_T$;
Stop.

## SUPPLEMENTARY MATERIAL

**AFLP data for sedges** (DOI: 10.1214/08-AOAS212SUPP). The data contains 126 markers from 2 plates for 14 species. The first column denotes the marker length. The names of these species are abbreviated as: Be (*Carex bebbii*), Bi (*C. bicknellii*, F (*C. festucacea*), N (*C. normalis*), O (*C. oronensis*), Te (*C. tenera* var. *echinodes*), Tt (*C. tenera* var. *tenera*) and Ti (*C. tincta*).

## REFERENCES

ALBERTSON, R. C., MARKERT, J. A., DANLEY, P. D. and KOCHER, T. D. (1996). Phylogeny of a rapidly evolving clade: The cichlid fishes of Lake Malawi, East Africa. *Proc. Natl. Acad. Sci. USA* **96** 5107–5110.
FELSENSTEIN, J. (1992). Phylogenies from restriction sites: A maximum-likelihood approach. *Evolution* **46** 159–173.
FELSENSTEIN, J. (2004). *Inferring Phylogenies* 251–256. Sinauer Associates, Inc., Sunderland, MA.
FLYBASE (2007). http://flybase.bio.indiana.edu. Downloaded in May, 2007.
GELMAN, A. and RUBIN, D. B. (1992). Inference from iterative simulation using multiple sequences. *Statist. Sci.* **7** 457–511.
GREEN, P. J. (1995). Reversible jump Markov chain Monte Carlo computation and Bayesian model determination. *Biometrika* **82** 711–732. MR1380810
GREEN, P. J. (2003). Trans-dimensional Markov chain Monte Carlo. In *Highly Structured Stochastic Systems* (P. J. Green, N. L. Hjort and S. Richardson, eds.) 179–198. Oxford Univ. Press, Oxford. MR2082410
HIPP, A. L., ROTHROCK, P. E., REZNICEK, A. A. and BERRY, P. E. (2006). Chromosome number changes associated with speciation in sedges: A phylogenetic study in Carex section Ovales (Cyperaceae) using AFLP data. In *Monocots: Comparative Biology and Evolution* (J. T. Columbus, E. A. Friar, J. M. Porter, L. M. Prince and M. G. Simpson, eds.). Rancho Santa Ana Botanic Garden, Claremont, CA.
HUELSENBECK, J. P. and RONQUIST, F. (2001). MRBAYES: Bayesian inference of phylogenetic trees. *Bioinformatics* **17** 754–755.
JONES, C. J., EDWARDS, K. J., CASTAGLIONE, S., WINFIELD, M. O., SALA, F., VAN DE WIEL, C., BREDEMEIJER, G., VOSMAN, B., MATTHES, M., DALY, A., BRETTSCHNEIDER, R., BETTINI, P., BUIATTI, M., MAESTRI, E., MALCEVSCHI, A., MARMIROLI, N., AERT, R., VOLCKAERT, G., RUEDA, J., LINACERO, R., VAZQUEZ, A. and KARP, A. (1997). Reproducibility testing of RAPD, AFLP and SSR markers in plants by a network of European laboratories. *Molecular Breeding* **3** 381–390.

Jukes, T. H. and Cantor, C. R. (1969). Evolution of protein molecules. In *Mammalian Protein Metabolism, Vol. 3* (H. N. Munro, ed.) 21–132. Academic Press, New York.

Luo, R. (2007). Bayesian study of AFLP marker evolution in phylogenetic inference. Ph.D. thesis, Dept. of statistics, Univ. Wisconsin–Madison.

Luo, R., Hipp, A. and Larget, B. (2007). A Bayesian model of AFLP marker evolution and phylogenetic inference. *Statist. Appl. Genet. Mol. Biol.* **6** Article 11. MR2306946

Luo, R. and Larget, B. (2009). Supplement to "Modeling substitution and indel processes for AFLP marker evolution and phylogenetic inference." DOI: 10.1214/08-AOAS212SUPP.

Mau, B. and Newton, M. A. (1997). Phylogenetic inference for binary data on dendograms using Markov chain Monte Carlo. *J. Comput. Graph. Statist.* **6** 122–131.

Miklós, I., Lunter, G. A. and Holmes, I. (2004). A "long indel" model for evolutionary sequence alignment. *Mol. Biol. Evol.* **21** 529–540.

NCBI (2007). http://www.ncbi.nlm.nih.gov/Genomes/. Downloaded in May, 2007.

Powell, W., Morgante, M., Andre, C., Hanafey, M., Vogel, J., Tingey, S. and Rafalski, A. (1996). The comparison of RFLP, RAPD, AFLP and SSR markers for germplasm analysis. *Molecular Breeding* **2** 225–238.

Redelings, B. D. and Suchard, M. A. (2005). Joint Bayesian estimation of alignment and phylogeny. *Systematic Biology* **54** 401–418.

Tamura, K. and Nei, M. (1993). Estimation of the number of nucleotide substitutions in the control region of mitochondrial DNA in humans and chimpanzees. *Mol. Biol. Evol.* **10** 512–526.

Thorne, J. L., Kishino, H. and Felsenstein, J. (1991). An evolutionary model for maximum likelihood alignment of DNA sequences. *Journal of Molecular Evolution* **33** 114–124.

Thorne, J. L., Kishino, H. and Felsenstein, J. (1992). Inching toward reality: An improved likelihood model of sequence evolution. *Journal of Molecular Evolution* **34** 3–16.

Tierney, L. (1994). Markov chains for exploring posterior distributions. *Ann. Statist.* **22** 1701–1728. MR1329166

Vos, P., Hogers, R., Bleeker, M., Reijans, M., van de Lee, T., Hornes, M., Frijters, A., Pot, J., Peleman, J., Kuiper, M. and Zabeau, M. (1995). AFLP: A new technique for DNA fingerprinting. *Nucleic Acids Research* **23** 4407–4414.

Wolfe, A. D. and Liston, A. (1998). *Molecular Systematics of Plants II* 43–86. Kluwer Academic, Boston, MA.

Yang, Z. (1993). Maximum likelihood estimation of phylogeny from DNA sequences when substitution rates differ over sites. *Mol. Biol. Evol.* **10** 1396–1401.

Yang, Z. (1994). Maximum likelihood estimation of phylogeny from DNA sequences with variable rates over sites: Approximate methods. *Journal of Computational Evolution* **39** 306–314.

Department of Statistics
Medical Science Center
University of Wisconsin—Madison
1300 University Avenue
Madison, Wisconsin 53706
USA
E-mail: rluo@stat.wisc.edu
       larget@stat.wisc.edu